\documentclass[12pt]{article}
\usepackage{amssymb}
\def\be{\begin{equation}}
\def\ee{\end{equation}}
\def\bes{\begin{eqnarray}}
\def\ees{\end{eqnarray}}
\def\p{\partial}

\def\half{{\textstyle{1\over2}}}

\catcode`\@=11
\def\@citex[#1]#2{%
\if@filesw \immediate \write \@auxout {\string \citation {#2}}\fi
\@tempcntb\m@ne \let\@h@ld\relax \def\@citea{}%
\@cite{%
  \@for \@citeb:=#2\do {%
    \@ifundefined {b@\@citeb}%
      {\@h@ld\@citea\@tempcntb\m@ne{\bf ?}%
      \@warning {Citation `\@citeb ' on page \thepage \space undefined}}%
      {\@tempcnta\@tempcntb \advance\@tempcnta\@ne%
      \@tempcntb\number\csname b@\@citeb \endcsname \relax%
      \ifnum\@tempcnta=\@tempcntb %Number follows previous--hold on to it
        \ifx\@h@ld\relax%
          \edef \@h@ld{\@citea\csname b@\@citeb\endcsname}%
        \else%
          \edef\@h@ld{\ifmmode{-}\else--\fi\csname b@\@citeb\endcsname}%
        \fi%
      \else%   %  non-successor--dump what's held and do this one
        \@h@ld\@citea\csname b@\@citeb \endcsname%
        \let\@h@ld\relax%
      \fi}%
    \def\@citea{,\penalty\@highpenalty\,}%
  }\@h@ld
}{#1}}

\def\@citeb#1#2{{[#1]\if@tempswa , #2\fi}}
\def\@citeu#1#2{{$^{#1}$\if@tempswa , #2\fi }}
\def\@citep#1#2{{#1\if@tempswa , #2\fi}}

\def\bcites{         % cite with []'s
        \catcode`\@=11
        \let\@cite=\@citeb
        \catcode`\@=12
}

\def\upcites{         % cite with exponents
        \catcode`\@=11
        \let\@cite=\@citeu
        \catcode`\@=12
}

\def\plaincites{      % cite without brackets
        \catcode`\@=11
        \let\@cite=\@citep
        \catcode`\@=12
}

\newcount\hour
\newcount\minute
\newtoks\amorpm
\hour=\time\divide\hour by 60
\minute=\time{\multiply\hour by 60 \global\advance\minute by-\hour}
\edef\standardtime{{\ifnum\hour<12 \global\amorpm={am}%
        \else\global\amorpm={pm}\advance\hour by-12 \fi
        \ifnum\hour=0 \hour=12 \fi
        \number\hour:\ifnum\minute<10 0\fi\number\minute\the\amorpm}}
\edef\militarytime{\number\hour:\ifnum\minute<10 0\fi\number\minute}

\def\draftlabel#1{{\@bsphack\if@filesw {\let\thepage\relax
   \xdef\@gtempa{\write\@auxout{\string
      \newlabel{#1}{{\@currentlabel}{\thepage}}}}}\@gtempa
   \if@nobreak \ifvmode\nobreak\fi\fi\fi\@esphack}
        \gdef\@eqnlabel{#1}}
\def\@eqnlabel{}
\def\@vacuum{}
\def\marginnote#1{}
\def\draftmarginnote#1{\marginpar{\raggedright\scriptsize\tt#1}}
\def\draft{
        \pagestyle{plain}
        \overfullrule=2pt
        \oddsidemargin -.5truein
        \def\@oddhead{\sl \phantom{\today\quad\militarytime} \hfil
        \smash{\Large\sl DRAFT} \hfil \today\quad\militarytime}
        \let\@evenhead\@oddhead
        \let\label=\draftlabel
        \let\marginnote=\draftmarginnote
        \def\ps@empty{\let\@mkboth\@gobbletwo
        \def\@oddfoot{\hfil \smash{\Large\sl DRAFT} \hfil}
        \let\@evenfoot\@oddhead}
        \def\@eqnnum{(\theequation)\rlap{\kern\marginparsep\tt\@eqnlabel}%
        \global\let\@eqnlabel\@vacuum}  }
\begin{document}

%\hfill CERN-TH/2001-310

\hfill UTHET-03-0801

%\hfill {\tt hep-th/yymmxxx} 
\vspace{-0.2cm}

\begin{center}
\Large
{\bf The Schwarzschild solution in the DGP model}
\normalsize

\vspace{0.8cm}

{\bf Chad Middleton\footnote{cmiddle1@utk.edu} and George 
Siopsis\footnote{gsiopsis@utk.edu}}\\ {\em Department of Physics
and Astronomy, \\
The University of Tennessee, Knoxville \\
TN 37996 - 1200, USA.} \\
June 2004
\end{center}

\vspace{0.8cm}

\centerline{\bf Abstract}

We discuss the Schwarzschild solution in the Dvali-Gabadadze-Porrati (DGP) model.  
We obtain a perturbative expansion and find the explicit form of the lowest-order contribution.
By keeping off-diagonal terms in the metric, we arrive at a perturbative expansion which is valid both far from and near the Schwarzschild radius.
We calculate the lowest-order contribution explicitly and obtain the form of
the metric both on the brane and in the bulk.
As we approach the Schwarzschild radius, the perturbative
expansion yields the standard four-dimensional Schwarzschild solution on the brane which is non-singular in the decoupling limit.
This non-singular behavior is similar to the Vainshtein solution in massive gravity demonstrating the absence of the van Dam-Veltman-Zakharov (vDVZ) discontinuity in the DGP model.

\newpage

Extra dimensions have been successful in explaining the weakness of gravity~\cite{bib1,bib2,bib3,bib4,bib5}.
They modify Newton's Law by introducing Kaluza-Klein modes.
In the case where the extra volume is infinite, light Kaluza-Klein modes dominate even at low energies~\cite{bib6,bib7,bib8},
leading to a modification of Newton's Law at astronomically large distances~\cite{bib5a,bib5b,bib5c,bib5d,bib5e,bib5f,bib5g,bib5h,bib5i}.
However, this may be avoided in the case of more than one extra dimension~\cite{bib7}.
Problems arise when the brane is given a finite thickness in the transverse
directions~\cite{bibkir,bibVVVV,bibus} obscuring the zero-thickness limit.
\footnote{See \cite{JR} for a consistent, finite-thickness brane model.}

The case of a four-dimensional brane living in a five-dimensional space
of vanishing cosmological constant (Dvali-Gabadadze-Porrati (DGP) model~\cite{bib6})
appears to be plagued by the van Dam-Veltman-Zakharov (vDVZ) discontinuity \cite{vanDam,Zak} and 
has recently attracted some attention~\cite{Deffayet,Lue,Gruzinov,Porrati,Lue1,bibkof,bibkof1,Tana}.
This is similar to four-dimensional massive gravity where one does not obtain agreement with Einstein's General Theory of Relativity in the limit of vanishing
mass of the graviton.
Vainstein~\cite{Vain} provided a solution to the apparent discontinuity in
the case of a point source by suggesting 
that the discrepancy arises from the linear approximation to the full 
field equations which has a limited range of validity.
%Below a certain critical radius, one had to perform a different perturbative expansion.
This second solution reduced to the
Schwarzschild solution in the zero graviton mass limit demonstrating the
absence of the vDVZ discontinuity in this spherically symmetric case.
\footnote{See~\cite{Papa} for problems associated with Vainstein's approach.}

Applying a similar procedure to the DGP model is not straightforward, because
the non-linear field equations are too complicated to solve even in the spherically symmetric case of
a point source.
%A solution was found in the case of a uniform source (cosmological solution)~\cite{Deffayet}.
It was subsequently argued in~\cite{Porrati}
that the discontinuity can be avoided by bending the brane through a coordinate
transformation. This bending can be large even if the matter sources are weak.
In refs.~\cite{Lue1,bibkof,bibkof1,Tana} solutions were found interpolating between the regimes
far from and near the Schwarzschild radius by keeping second-order bulk derivatives in the perturbative expansion.
It was thus shown that in the decoupling limit one recovers the standard four-dimensional Schwarzschild metric. However, the method of solution in~\cite{Lue1,bibkof,bibkof1,Tana} only yielded the form of the metric on the brane and
provided no information on the bulk dependence of the metric.

Here we discuss a perturbative solution to the DGP field equations in the case of a point source.
By employing an {\em ansatz} for the metric containing off-diagonal contributions,
we arrive at a lowest-order approximation to the field equations which is
solvable.
The solution is found explicitly throughout its domain of validity (both near and far from the Schwarzschild radius), including the bulk.

The DGP model~\cite{bib6} describes a 3-brane on the boundary of a five-dimensional bulk space $\Sigma$. The action is  
\be\label{eq2}
S=M^{3}\int_\Sigma d^4xdy\sqrt{-G}\mathcal{R}_{(5)} + \bar M^2
\int_{\p\Sigma} d^4x\sqrt{-\overline G}\; \overline\mathcal{R}_{(4)} 
\ee
where $\mathcal{R}_{(5)}\;(\overline\mathcal{R}_{(4)})$ is the five-(four-)
dimensional 
Ricci Scalar.
The solution to the linearized equations bares a striking 
resemblence to the vDVZ solution of massive gravity and 
shares the apparent discontinuity in the decoupling limit 
$(M\rightarrow 0)$.  Porrati~\cite{Porrati} argued that this
solution is only valid in a limited domain, as was the vDVZ solution of 
massive gravity, and breaks down in the regime
\be\label{eq26}
r\lesssim r_c \ \ , \ \ r_c = \left(\frac{m\bar{M}^2}{18\pi M^6}\right)^{1/3}
\ee
when a static spherically-symmetric source of mass $m$ is considered. 

We seek to obtain a solution to the field equations for a static point source
with stress-energy tensor
\be T_{AB} = m \delta_A^0\delta_B^0 \delta^3(\vec x)\delta(y)\ee
which is valid throughout the region $r_m \lesssim r \lesssim 1/m_b$, where
\be r_m= \frac{2m}{8\pi\bar{M}^2}\ee
is the Schwarzschild radius and
\be m_b = \frac{2M^3}{\bar M^2}\ee
is a crossover scale between four-dimensional and five-dimensional behavior.
In the decoupling limit, $m_b\to 0$ and $r$ is unbounded from above.

We choose the {\em ansatz} for the metric
\be
ds^2=-e^{2B(r,y)}dt^2+e^{2C(r,y)}\delta_{ij}dx^idx^j+2A_i(r,y)dx^idy+e^{2D(r,y)}dy^2
\ee
To arrive at a set of compatible field equations, we shall keep first-order
contributions in the diagonal components $B,C,D$, and up to {\em second-order}
terms in the off-diagonal field $\vec A$.
%it will be necessary to assume that $B,C,D$ are of 
%first-order ($o(m)$), whereas $\phi$ is of  {\em half}-order ($\phi\sim o(\sqrt{m})$).
It is also convenient to introduce the notation
\be\label{eqApsi}
\vec{A}=\vec{\nabla}\phi\:\:,\:\:\Psi'=\frac{1}{r}(\phi')^2
\ee
where prime denotes differentiation with respect to $r$, the distance from the point source.
A dot will be used for bulk derivatives (with respect to $y$).

First, let us discuss the lowest order contributions to the field equations in the bulk.
The $yy$ component reads
%\be
%R_y^y=R_\alpha^\alpha
%\ee
%
\be\label{eqyy}
2\p_i\p^i(B+2C)+\p_j(\p^j\phi\p_i\p^i\phi)-\p_j(\p^i\phi\p_i\p^j\phi) =0
\ee
The mixed components are
\be\label{eqi51}
\p_i(\dot{B}+2\dot{C})-\p^j\phi\p_j\p_i\dot{\phi} =0
\ee
The spatial brane worldvolume components are
$$
(\p_i\p_j-\delta_{ij}\p_k\p^k)(B+C+D-\dot{\phi})-\delta_{ij}(\ddot{B}+2\ddot{C})
$$
\be\label{eq10}
+ \p_k(\p^k\phi\p_i\p_j\phi)-\p_j(\p^k\phi\p_i\p_k\phi)-\half 
\delta_{ij}\left(\p_k(\p^k\phi\p_l\p^l\phi)-\p_k(\p^l\phi\p_l\p^k\phi)\right) =0
\ee
and finally, the $tt$ component is
\be\label{eq11}
\p_i\p^i(2C+D-\dot{\phi})+3\ddot{C}
+\half 
\left[\p_k(\p^k\phi\p_i\p^i\phi)-\p_j(\p^k\phi\p_k\p^j\phi)\right] =0
\ee
In terms of the field $\Psi$ (eq.~(\ref{eqApsi})), the field equation~(\ref{eqyy}) becomes linear, 
\be\label{eq55}
B+2C+\Psi=0
\ee
Then the mixed components~(\ref{eqi51}) may be written as
\be\label{eqi51a}
\p_i\dot\Psi +\p^j\phi\p_j\p_i\dot{\phi} =0
\ee
whose general solution is
\be \phi' (r,y) = \frac{\alpha(y)}{r^2} + \beta(r)\ee
We shall make use of gauge freedom and demand $\phi$
(as well as $\Psi$) be independent of $y$,
%The above two equations are satisfied if we choose 
\be\label{eqi5}
\dot{\phi}= \dot\Psi = 0 
\ee
This is true to lowest order; higher-order corrections will introduce a non-vanishing $\dot{\phi}$ (and $\dot\Psi$).
% Notice that in the DGP solution discussed in the previous section, we obtained $B+2C =0$.
%This conclusion is avoided in this case thanks to the quadratic term in $\phi$
%in eq.~(\ref{eqi51}).

The remaining field equations~(\ref{eq10}) and (\ref{eq11}) also become linear.
They read, respectively,
\be\label{eqij}
B+C+D+\half\Psi =0
\ee
\be\label{eqtt}
\nabla^2 (2C+D+\Psi)+3\ddot{C}=0
\ee
where we used~(\ref{eqi5}).

Eqs.~(\ref{eq55}), (\ref{eqi5}) and (\ref{eqij}) yield
\be\label{eqBCD} 
C = -\half (B+\Psi) \ \ , \ \ D= -\half B \ee
Then eq.~(\ref{eqtt}) becomes
\be \ddot B + \nabla^2 B = 0\ee
whose solution is easily obtained after Fourier-transforming the worldvolume coordinates
\be\label{eq19} \tilde B(p, y) = \tilde B(p, 0) e^{-py}\ee
The bulk behavior of the other fields is then found from~(\ref{eqBCD}),
\be\label{eq19a} \tilde C(p,y) = -\half \tilde B(p,0) e^{-py} -\half \tilde\Psi (p)
\ \ , \ \
\tilde D(p,y) = -\half \tilde B(p,0) e^{-py}\ee
Having obtained the functional dependence of all fields on $y$ in terms of
data on the brane ($y=0$), we now turn to solving the boundary field equations.

On the boundary ($y=0$), the spatial brane components yield
\be\label{eqijb}
2M^3\phi+\bar{M}^2(B + C) = 0
\ee
whereas the $tt$ component is
\be\label{eqttb}
6M^3 \dot{C} + 2\bar{M}^2\p_i\p^iC+2M^3\p_i\p^i\phi=-m\delta^3(\vec{x})
\ee
The first term in eq.~(\ref{eqttb}) may be dropped in the regime of interest,
$p\gg m_b = 2M^3/\bar M^2$.
Eliminating $C$ by using~(\ref{eqBCD}), we obtain
\be
\bar{M}^2\nabla^2 (B+\Psi) -2M^3\nabla^2 \phi
= m\delta^3 (\vec x)
\ee
Solving for $B$ on the boundary, we find
\be\label{eqB0}
 B(r,0) = -\Psi + m_b\phi - \frac{m}{4\pi \bar M^2 r}
\ee
Using this and (\ref{eqBCD}) to eliminate $B$ and $C$ from the other boundary eq.~(\ref{eqijb}), we deduce
\be\label{eqphi}
\frac{3}{2} m_b \phi - \Psi - \frac{m}{8\pi \bar M^2 r} =0
\ee
Differentiating with respect to $r$ and using (\ref{eqApsi}),
\be 
\frac{3}{2} m_b \phi' - \frac{1}{r}(\phi')^2 + \frac{m}{8\pi \bar M^2 r^2} =0 \ee
which can easily be solved for $\phi'$ yielding,
\be\label{eqphip}
\phi' = \frac{3}{4} m_br \left(1 - \sqrt{1+\frac{r_c^3}{r^3}} \right)
\ee
where $r_c$ is given by eq.~(\ref{eq26}).
We also have
\be\label{eqpsip} \Psi' = \frac{1}{r} (\phi')^2 = \frac{m}{8\pi\bar{M}^2r^2} +\frac{9}{8}m_b^2r\left(1 - \sqrt{1+\frac{r_c^3}{r^3}} \right)\ee
Differentiating eq.(\ref{eqB0}) with respect to $r$, we obtain
the form of the field $B$ on the brane,
\be\label{eqBC}
B'(r,0) = \frac{m}{8\pi\bar{M}^2r^2}-\frac{3}{8}m_b^2r\left(1 - \sqrt{1+\frac{r_c^3}{r^3}} \right)
\ee
Summarizing, eqs.~(\ref{eqBCD}), (\ref{eqphip}) and (\ref{eqBC}) provide the
form of the metric on the brane.
This solution is valid everywhere on the brane (in the regime $r_m\lesssim r
\lesssim 1/m_b$).  The solution in the bulk is given by eqs.~(\ref{eq19})
and (\ref{eq19a}) where the fourier transform $\tilde B(p,0)$
is deduced from~(\ref{eqBC}).

Next, we examine the near and far regimes, separated by the crossover distance $r_c$ (eq.~(\ref{eq26})). 

In the far regime ($r\gtrsim r_c$), we have from~(\ref{eqphip}),
\be 
\phi= \frac{m}{24\pi M^3 r}\left\{1+o(r_c^3/r^3)\right\}
\ee
and so (using eq.~(\ref{eqApsi}) or (\ref{eqpsip})),
\be\Psi = - \frac{m}{128\pi \bar M^2 r}\ \left\{ \frac{r_c^3}{r^3}+o(r_c^6/r^6)\right\}\ee
Notice that $\Psi$ is of higher order. Therefore, eq.~(\ref{eqBCD}) implies to
lowest order
\be C(r,y) = D(r,y) = -\half B(r,y)\ee
Using (\ref{eqBC}), we obtain
\be
 B(r,0)= -\frac{m}{6\pi \bar M^2 r}\ \left\{ 1+o(r_c^3/r^3)\right\}
\ee
and after fourier transforming,
\be \tilde B(p,0) = - \frac{2m}{3\bar M^2 p^2}\ \left\{ 1+o(p^3/p_c^3)\right\}\ee
where $p_c \sim 1/r_c$.
The $y$ dependence of the fields to lowest order is given by
\be
\tilde{C}(p,y) =\tilde D(p,y) = -\half \tilde{B}(p,y) = \frac{m}{3\bar{M}^2p^2}e^{-py} 
\ee
Now taking the inverse fourier transforms, we finally obtain for $r\gtrsim r_c$,
\bes
 B_> (r,y) &=& -\frac{m}{6\pi \overline M^2 r}\ \left\{ 1 - \frac{2}{\pi} \tan^{-1}
\frac{y}{r} \right\} \left( 1 + o (r_c^3/r^3)\right)\nonumber\\
 C_> (r,y) &=& \frac{m}{12\pi \overline M^2 r}\ \left\{ 1 - \frac{2}{\pi} \tan^{-1}
\frac{y}{r} \right\} \left( 1 + o (r_c^3/r^3)\right)\nonumber\\
 D_> (r,y) &=& \frac{m}{12\pi \overline M^2 r}\ \left\{ 1 - \frac{2}{\pi} \tan^{-1}
\frac{y}{r} \right\} \left( 1 + o (r_c^3/r^3)\right)\nonumber\\
\phi_> (r,y) &=& \frac{m}{24\pi M^3 r}\left( 1+o(r_c^3/r^3)\right)
\ees
This solution corresponds to that of the standard perturbative expansion.  Notice that in the decoupling limit ($M^3\rightarrow 0$), $\phi$ diverges. But this
is a limit beyond our approximation above, because corrections become infinite
($r_c\to\infty$, from eq.~(\ref{eq26})). We next obtain the solution in the near regime which corresponds to the four-dimensional Schwarzschild solution and is valid in the decoupling limit.

In the near regime ($r\lesssim r_c$), we obtain from eqs.~(\ref{eqphip}) and
(\ref{eqpsip}), respectively,
\be\label{eq38}
 \phi = - \sqrt{\frac{mr}{2\pi \bar M^2}}\ \left\{ 1 + o ((r/r_c)^{3/2})\right\}
\ \ , \ \ \Psi = -\frac{m}{8\pi \bar M^2 r}\ \left\{ 1 + o ((r/r_c)^{3/2})\right\}
\ee
In this case $\Psi$ contributes to lowest order.
Using (\ref{eqBC}), we deduce on the brane
\be
\tilde B(p,0)=- \frac{m}{2 \bar M^2 p^2}\ \left\{ 1 + o ((p_c/p)^{3/2})\right\}
\ee
The $y$-dependence to lowest order is given by
\bes\label{eq40}
\tilde{B}(p,y) &=& -\frac{m}{2\bar{M}^2p^2}e^{-py}\nonumber\\
\tilde{C}(p,y) &=& \frac{m}{4\bar{M}^2p^2}(1+e^{-py})\nonumber\\
\tilde{D}(p,y) &=& \frac{m}{4\bar{M}^2p^2}e^{-py}
\ees
where we used~(\ref{eq19}) and (\ref{eq19a}).
At $y=0$, we recover the Schwarzschild solution and therefore agreement with the standard Newtonian potential of massless gravity.  All fields are now non-singular in the decoupling limit $M\rightarrow 0$.
Fourier transforming, in the regime $r_m \lesssim r \lesssim r_c$, we obtain
from eqs.~(\ref{eq38}) and (\ref{eq40}),
\bes
 B_< (r,y) &=& -\frac{m}{8\pi \overline M^2 r}\ \left\{ 1 - \frac{2}{\pi} \tan^{-1}
\frac{y}{r} \right\} \left( 1 + o ((r/r_c)^{3/2})\right)\nonumber\\
 C_< (r,y) &=& \;\;\frac{m}{8\pi \overline M^2 r}\ \left\{ 1 - \frac{1}{\pi} \tan^{-1}
\frac{y}{r} \right\} \left( 1 + o ((r/r_c)^{3/2})\right)\nonumber\\
 D_< (r,y) &=& \frac{m}{16\pi \overline M^2 r}\ \left\{ 1 - \frac{2}{\pi} \tan^{-1}
\frac{y}{r} \right\} \left( 1 + o ((r/r_c)^{3/2})\right)\nonumber\\
 \phi_< (r,y) &=& - \sqrt{\frac{mr}{2\pi \bar M^2}}\ \left( 1 + o ((r/r_c)^{3/2})\right)
\ees
%We discussed the Schwarzschild solution in the Dvali-Gabadadze-Porrati (DGP) model.
%We derived two different perturbative expansions, (\ref{eq1st}) and (\ref{eq2nd}), which were valid in two distinct domains, $r\gtrsim r_c$ and $r\lesssim r_c$, respectively.
%We showed that the critical radius $r_c$~\cite{Porrati} can be determined by the lowest-order contributions to the two perturbative expansions using the criterion~(\ref{eqcr}).
To conclude, we derived a perturbative expansion for the Schwarzschild solution in the DGP model.
By keeping terms in the field equations which were second-order in the off-diagonal metric components,
we arrived at an explicit form of the solution both on the brane and in the bulk.
On the brane, our solution interpolated between the near and far regimes separated by the distance scale $r_c$ (eq.~(\ref{eq26})).
At distances below the critical radius $r_c$, the perturbative
expansion yielded the four-dimensional Schwarzschild solution, demonstrating the absence of the van Dam-Veltman-Zakharov (vDVZ) discontinuity~\cite{vanDam,Zak}.

\vspace{0.5cm}

\section*{Acknowledgements}

We thank G.~Kofinas, A.~Lue, A.~Papazoglou and J.~Rombouts for useful discussions and insightful comments.  This work is  
supported in part by the DoE under grant DE-FG05-91ER40627.

\end{document}